\documentclass[showpacs,superscriptaddress,amsmath,aps]{revtex4}
\usepackage{graphicx,amssymb,longtable,mathrsfs}

\newcommand{\nl}{\nonumber \\}
\newcommand{\ie}{{\it i.e.}}

\newcommand{\be}{\begin{equation*}}
\newcommand{\ee}{\end{equation*}}
\newcommand{\bea}{\begin{eqnarray}}
\newcommand{\eea}{\end{eqnarray}}
\newcommand{\bmp}{\begin{minipage}}
\newcommand{\emp}{\end{minipage}}
\newcommand{\pb}{\parbox}

\newcommand{\Eq}[1]{Eq.\,(\ref{#1})}

\newcommand{\alf}{\alpha}
\newcommand{\sgm}{\sigma}
\newcommand{\omg}{\omega}
\newcommand{\Omg}{\Omega}
\newcommand{\vpl}{\varepsilon}
\newcommand{\wit}{\widetilde}
\newcommand{\Gam}{\Gamma}
\newcommand{\upa}{\uparrow}
\newcommand{\dwa}{\downarrow}
\newcommand{\la}{\langle}
\newcommand{\ra}{\rangle}
\newcommand{\dg}{\dagger}

\newcommand{\mb}{\mbox}

\begin{document}
\draft

\title{ Calculation of the current noise spectrum in mesoscopic transport:
        an efficient quantum master equation approach }

\author{JunYan Luo}
\affiliation{State Key Laboratory for Superlattices and Microstructures,
         Institute of Semiconductors,
         Chinese Academy of Sciences, P.O.~Box 912, Beijing 100083, China}
\author{Xin-Qi Li}
\email{xqli@red.semi.ac.cn}
\affiliation{State Key Laboratory for Superlattices and Microstructures,
         Institute of Semiconductors,
         Chinese Academy of Sciences, P.O.~Box 912, Beijing 100083, China}
\author{YiJing Yan}
\affiliation{Department of Chemistry, Hong Kong University of Science and
         Technology, Kowloon, Hong Kong}

\date{\today}

\begin{abstract}
Based on our recent work on quantum transport
[Li {\it et al.}, Phys. Rev. B {\bf 71}, 205304 (2005)], where the calculation of
transport current by means of quantum master equation was presented,
in this paper we show how an efficient calculation can be performed
for the transport noise spectrum.
Compared to the longstanding classical rate equation or
the recently proposed quantum trajectory method,
the approach presented in this paper combines their respective advantages,
i.e., it enables us to tackle both the many-body Coulomb interaction
and quantum coherence on equal footing
and under a wide range of setup circumstances.
The practical performance and advantages are illustrated by a number of examples,
where besides the known results and new insights obtained in a transparent manner,
we find that this alternative
approach is much simpler than other well-known full quantum mechanical methods
such as the Landauer-B\"uttiker scattering matrix theory and the nonequilibrium
Green's function technique.
\\
\end{abstract}
\maketitle


\section{\label{sec:1}Introduction}

A modern trend in transport studies of mesoscopic systems is not only for
the current-voltage characteristics, but also for the noise properties
\cite{Land,Loss,Schol,Schon,Cottet,Ciam,Birk,Safo,Nauen,Jung}.
The noise spectrum, which is a measure of the temporal correlation
between individual electron events,
has been proved to be a unique tool to reveal different possible mechanisms
which are not accessible by the mean current measurement \cite{Butti}.
In particular, via the measurement of current noise
one is also able to extract information of the system parameters that govern the
transport, as well as the internal energy scales of the mesoscopic system.

In principle, full quantum mechanical calculation for transport noise spectrum can be
performed by either the Landauer-B\"uttiker scattering matrix approach
\cite{Butti91,Butti} or the nonequilibrium Green's function (nGF) technique
\cite{Ch-T,Hung}. In practice, however, they were largely restricted to non-interacting
systems. Very recently, the {\it zero-frequency} noise of transport through quantum dot
with strong Coulomb interaction was calculated \cite{Schon03}, with the nGF-based
sophisticated real-time diagrammatic technique \cite{Schon96}.

An alternative method to calculate transport noise is the {\it classical} rate
equation approach \cite{Jong,Davies,Korotk}. This approach is much simpler than the full
quantum mechanical methods, thus has been employed in some interesting transport systems,
for instance, the important Coulomb blockade systems \cite{Wilk93,Hanke}.
Nevertheless, owing to the classical nature of this approach,
the {\it quantum coherence} that widely exists inside the transport systems
cannot be described.
Focused on the effect of quantum coherence in noise spectrum, Sun and Milburn
studied the transport through a pair of coupled quantum dots \cite{Milb}.
For this specific system, they derived a Lindblad-type master equation,
then unravelled it to calculate the noise spectrum.
With the same motivation to include quantum coherence,
an alternative ``$n$"-resolved quantum Bloch equation approach
was proposed some time earlier by Gurvitz \textit{et al.} \cite{Gurv96},
and recently applied it further \cite{Gurv05}.
(Here ``$n$" stands for the electron number tunnelled through the junctions).
We noticed that this approach started with the many-particle Schr\"odinger
equation of the entire system,
thus its applicability is restricted to {\it zero temperature}.
Also, the derivation limited its validity condition in {\it large bias voltage} regime.
In our recent work \cite{Li05,Li04,Li05-1}, we extended this approach to
finite temperature and arbitrary voltages, where we have
based our derivation on the quantum master equation, which is free
from any specific system, and can serve as a general and convenient
starting point to study quantum transport.

In Ref.\ \onlinecite{Li05-1}, we have established an explicit and compact expression
for transport current and demonstrated its application by a number of typical examples.
In this paper, we complete that master-equation transport formalism,
by developing further the formulation for noise spectrum calculation.
In next section, for completeness, we first outline the general idea of the
``$n$"-resolved transport master equation, then present the compact formulas
for transport current and noise spectrum.
In Sec. \ref{sec:3} we demonstrate the noise formula by the same examples in
Ref.\ \onlinecite{Li05-1}, i.e., transport through a non-interacting and interacting
quantum dot, and a pair of coherently coupled quantum dots.
For the simplest system of non-interacting dot, we easily recover the known result
obtained first by Chen and Ting, which was based on a full quantum mechanical
treatment in terms of non-equilibrium path-integral approach
\cite{Ch-T}.
For interacting dot, we extend the result of zero-frequency noise obtained
in Ref.\ \onlinecite{Schon03} to the whole frequency regime, and present
brief discussion for the noise characteristics.
For the coupled quantum dots, where the quantum coherence plays significant role,
we show that our approach can be as efficient as the quantum trajectory method \cite{Milb}.
However, by noting that the quantum trajectory theory is based on the unravelling
of Lindblad-type master equation, we believe that the formalism presented in this
work is more powerful than the quantum trajectory approach, since not all the transport
systems can be reduced to a Lindblad-type master equation
(e.g. the above mentioned transport through interacting quantum dot).
Finally, a brief conclusion is given in Sec. \ref{sectionconclusion}.

\section{\label{sec:2} FORMALISM}


Consider the typical transport setup as schematically shown in Fig. 1,
which consists of the central transport system (device) and two biased electrodes.
The total system Hamiltonian reads
\begin{eqnarray}\label{sysHamilton}
H&=&H_{S}(c^{\dag}_{\mu},c_{\mu})+\sum_{\alf=L,R} \sum_{k}
E_{\alf k}a^{\dag}_{\alf k}a_{\alf k}+\sum_{\alf=L,R}
\sum_{k\mu}(t_{\alf k \mu}a^{\dag}_{\alf k}c_{\mu}+\mb{H.c.}).
\end{eqnarray}
$H_S$ is the central system Hamiltonian, in which all the possible
many-body interactions have been included.
$c_\mu^\dag$ ($c_\mu$) is the electron creation (annihilation) operator
of state ``$\mu$", which labels here both the orbital and spin states.
The second term describes the non-interacting electrons of the two electrodes,
and the third one describes tunneling between the central system and the electrodes.

\begin{figure}
\begin{center}
\includegraphics*[width=7cm,height=5cm,keepaspectratio]{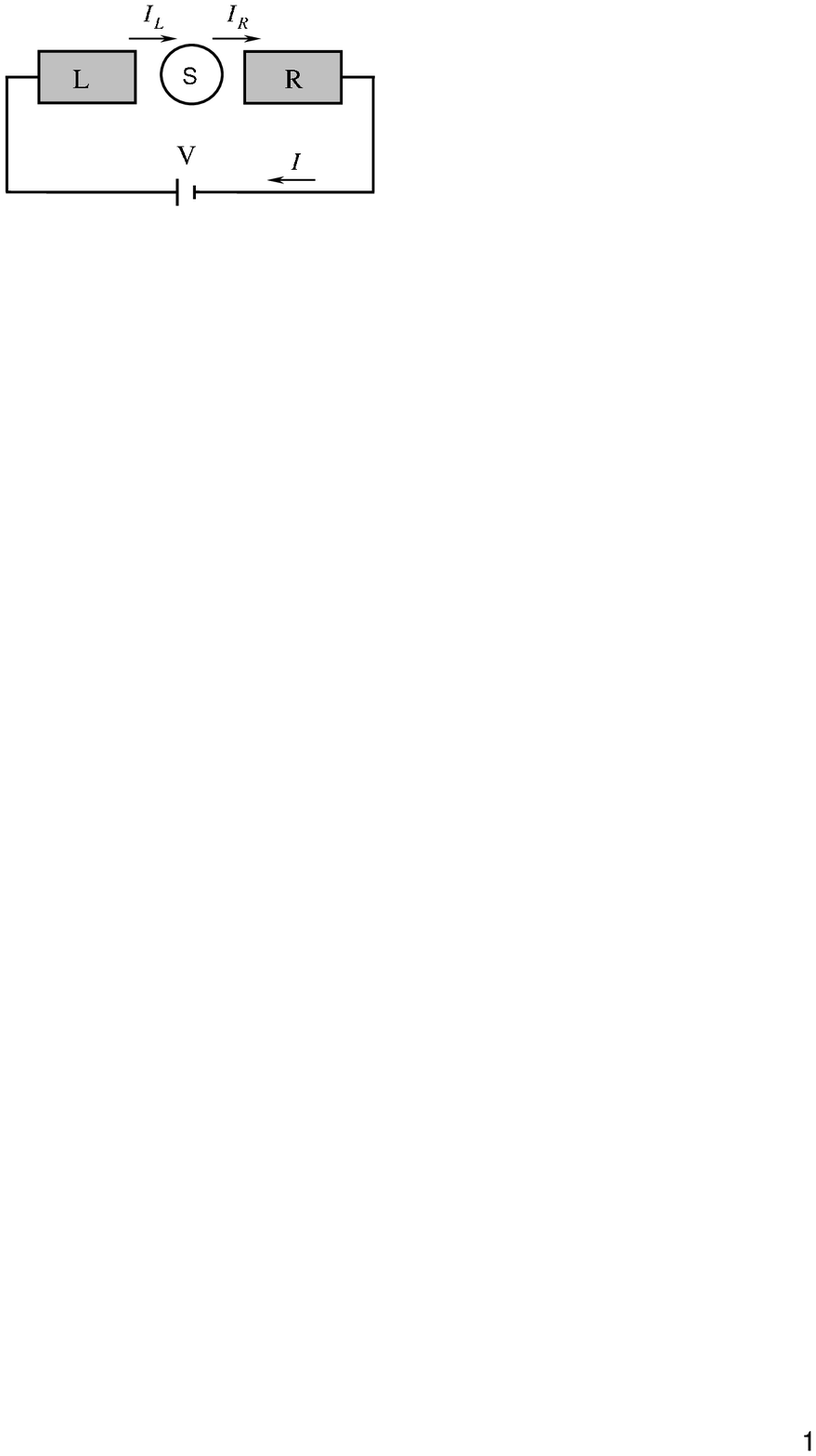}
\caption{Schematic setup for electron transport through a mesoscopic system. }
\end{center}
\end{figure}

For convenience, we re-express the tunneling Hamiltonian as
\begin{eqnarray}
H'=\sum_\mu \left(c^{\dag}_\mu F_\mu+\mb{H.c}\right),
\end{eqnarray}
where $F_\mu=F_{L\mu}+F_{R\mu}\equiv\sum_{\alf=L,R}t_{\alf k\mu}a_{\alf\mu\sgm}$.
By regarding the tunneling Hamiltonian as perturbation,
the second-order cumulant expansion leads to the master equation
for the reduced density matrix of the central transport system \cite{Yan}
\begin{eqnarray}\label{cumm-expan}
\dot{\rho}(t)=-i\mathcal{L} \rho(t)-\int_0^\infty
d\tau\la\mathcal{L}'(t)\mathcal{G}(t,\tau)\mathcal{L}'(\tau)
\mathcal{G}^\dag(t,\tau)\ra\rho(t).
\end{eqnarray}
Here the reduced density matrix $\rho(t)$ is introduced by tracing over the
electrode states from the density matrix $\rho_T$ of the entire system,
\ie, $\rho(t)\equiv \mb{Tr}_B[\rho_T(t)]$,
and $\la\cdots\ra\equiv\mb{Tr}_B[(\cdots)\rho_B]$,
with $\rho_B$ the density operator of the electrodes.
The Liouvillian superoperators are defined as
$\mathcal{L}A\equiv [H_S, A]$, $\mathcal{L}'A\equiv[H',A]$,
and $\mathcal{G}(t,\tau)A\equiv G(t,\tau)AG^\dag(t,\tau)$,
where $G(t,\tau)$ is the usual propagator (Green's function) associated
with only the system Hamiltonian $H_S$.

The trace in \Eq{cumm-expan} is over all the electrode degrees of freedom,
leading thus to the equation of motion of the {\it unconditional}
reduced density matrix of the system.
To describe the transport problem, we should keep track of the
record of electron numbers that have tunnelled through the right (left) junction,
in order to calculate the corresponding current through it.
To this end, the average over states in the entire Hilbert space ``$B$"
in \Eq{cumm-expan} should be replaced with states in the subspace ``$B^{(n)}$",
which corresponds to $n$-electrons
tunnelled through the right or left junction.
Let us take the right junction as an example. The trace of \Eq{cumm-expan}
over ``$B^{(n_R)}$" leads to a master-type equation for the {\it conditional}
reduced state $\rho^{(n_R)}(t)$ \cite{Li04}:
\begin{eqnarray}\label{CQMER}
\dot{\rho}^{(n_R)}&=&-i\mathcal{L}\rho^{(n_R)}
-\frac{1}{2}\sum_{\mu}\{[c^{\dag}_{\mu}A^{(-)}_{\mu}\rho^{(n_R)}
+\rho^{(n_R)}A^{(+)}_{\mu}c^{\dag}_{\mu}\nonumber\\
&&-A^{(-)}_{L\mu}\rho^{(n_R)}c^{\dag}_{\mu}
-c^{\dag}_{\mu}\rho^{(n_R)}A^{(+)}_{L\mu}-A^{(-)}_{R\mu}\rho^{(n_R-1)}c^{\dag}_{\mu}
-c^{\dag}_{\mu}\rho^{(n_R+1)}A^{(+)}_{R\mu}]+\mb{H.c.}\}.
\end{eqnarray}
Here $A_{\alpha\mu}^{(\pm)}=\sum_{\nu}
C_{\alpha\mu\nu}^{(\pm)}(\pm {\cal L})a_{\nu}$,
and $A_{\mu}^{(\pm)}=\sum_{\alpha=L,R}A_{\alpha\mu}^{(\pm)}$.
The spectral functions $C_{\alpha\mu\nu}^{(\pm)}(\pm {\cal L})$
are defined in terms of the Fourier transform of the reservoir
correlation functions, i.e.,
$ C_{\alpha\mu\nu}^{(\pm)}(\pm {\cal L})=\int^{\infty}_{-\infty} dt
  C_{\alpha\mu\nu}^{(\pm)}(t) e^{\pm i{\cal L}t}$.
The correlation functions in time domain are defined by
$\la f_{\alpha\mu}^{\dg}(t)f_{\alpha\nu}(\tau)\ra =
C_{\alpha\mu\nu}^{(+)}(t-\tau)$, and $\la
f_{\alpha\mu}(t)f^{\dg}_{\alpha\nu}(\tau)\ra =
C_{\alpha\mu\nu}^{(-)}(t-\tau)$,
where $\la \cdots \ra$ stands for $\mb{Tr}_B [(\cdots)\rho_B^{(0)}]$,
having the usual meaning of thermal average.
For $\rho^{(n_L)}(t)$, which describes the reduced system state
conditioned by the electron numbers tunnelled through the left junction,
a similar equation as Eq. (\ref{CQMER}) can be obtained by interchanging
the indices $``L"$ and $``R"$.
With the knowledge of the above {\it conditional} state $\rho^{(n_{L/R})}(t)$,
we are ready to calculate all the transport properties, such as the transport current
and the noise spectrum.


\subsection{Transport Current}

Throughout this paper, we assume the unit system of $\hbar=e=1$.
Straightforwardly, the current through the $\alf_{\rm th}$ (i.e. $L~\mb{or}~R)$
junction can be calculated via
$I_\alf(t)= d\la N_{\alf}(t)\ra/dt
=\sum_{n_\alf}n_\alf{\rm Tr}[\dot{\rho}^{(n_\alf)}(t)]$,
where ${\rm Tr}(\cdots)$ means trace over the internal states of the central system.
Obviously, ${\rm Tr}[\rho^{(n_\alf)}(t)]\equiv P(n_{\alpha},t)$ is nothing
but the probability by that there are
``$n_{\alpha}$" electrons tunnelled through the $\alpha_{\rm th}$ junction
until time ``$t$".
Simple algebra leads to \cite{Li04}
\begin{subequations}\label{current}
\begin{eqnarray}
I_\alf(t)&=&\mb{Tr}\big[\mathscr{T}^{(-)}_\alf\rho(t)+\mb{H.c.}\big],\\
\mathscr{T}^{(-)}_\alf\rho(t)&=&\frac{1}{2}\sum_\mu\big[A_{\alf\mu}^{(-)}\rho(t)c_\mu^\dag
-c_\mu^\dag\rho(t)A_{\alf\mu}^{(+)}\big]\label{Tnegalf},
\end{eqnarray}
\end{subequations}
where the unconditional density matrix $\rho(t)$, which is the sum of all
$\rho^{(n_{\alpha})}(t)$, satisfies a simple unconditional master equation
\begin{subequations}\label{QME}
\begin{eqnarray}
\dot{\rho}(t)&=&-i\mathcal{L}\rho(t)-[\mathcal{R}\rho(t)+\mb{H.c.}],\\
\mathcal{R}\rho(t)&=&\frac{1}{2}\sum_\mu\big[c_\mu^\dag,
A_{\mu}^{(-)}\rho(t)-\rho(t)A_{\mu}^{(+)}\big].
\end{eqnarray}
\end{subequations}

\subsection{Noise Spectrum}

Physically, the noise spectrum characterizes the temporal fluctuations of the entire
circuit current.
In steady state, the {\it average currents} through the left and right junctions are
equal to each other, but the temporal (i.e. time-dependent) fluctuating currents are not.
In general, the circuit current, which is typically the measured quantity in most
experiments, is a superposition of the left and right currents, i.e.,
$I(t)=aI_L(t)+bI_R(t)$.
Here the coefficients $a$ and $b$ depend on the symmetry of the transport setup
(e.g. the junction capacitances) \cite{Butti}, and satisfy $a+b=1$.
In combination with the charge conservation law, say, $I_L=I_R+\dot{Q}$,
with $Q$ the charge in the central system, we obtain
\begin{eqnarray}
I(t)I(0)=aI_L(t)I_L(0)+bI_R(t)I_R(0)-ab\dot{Q}(t)\dot{Q}(0).
\end{eqnarray}
Accordingly, the noise spectrum is a sum of three parts
\begin{eqnarray}\label{Sc}
S(\omg)=a S_L(\omg)+bS_R(\omg)-ab\omg^2S_Q(\omg),
\end{eqnarray}
where $S_{L/R}(\omg)$ is the noise spectrum of the current through
the left (right) junction,
and $S_Q(\omg)$ characterizes the charge fluctuations in the central system.
In the following we first develop general formulation for their calculation,
then demonstrate a number of typical examples.

For $S_{L/R}(\omg)$, we employ the MacDonald's formula \cite{MacD}
\begin{eqnarray}\label{MacD}
S_\alf(\omg)=2\,\omg\int_{0}^{\infty}dt\sin(\omg t) \frac{d}{dt}\la n_\alf^{2}(t)\ra,
\end{eqnarray}
where $\la n_\alf^{2}(t)\ra=\sum_{n_\alf}n_\alf^{2}[\mb{Tr}\rho^{(n_\alf)}(t)]
=\sum_{n_\alf}n_\alf^{2}P(n_{\alpha},t)$.
With the help of Eq. (\ref{CQMER}), we can show that
\bea \label{dn2dt}
\frac{d}{dt}\la n_\alf^{2}(t)\ra&=&\mb{Tr}\big[2\mathscr{T}^{(-)}_\alf N^\alf(t)
+\mathscr{T}^{(+)}_\alf\rho+\mb{H.c.}\big],
\eea
in which we introduce the ``particle-number" matrix
$N^\alf(t)\equiv\sum_{n_\alf}n_\alf\rho^{(n_\alf)}(t)$, and the superoperator means
\bea
\mathscr{T}^{(\pm)}_\alf(\cdots)&=&\frac{1}{2}\sum_\mu\big[A_{\alf\mu}^{(-)}(\cdots)
c_\mu^\dag \pm c_\mu^\dag(\cdots)A_{\alf\mu}^{(+)}\big].
\eea
Then, $S_{\alpha}(\omg)$ is formally expressed as
\begin{eqnarray}\label{Salpha}
S_\alf(\omg)=2\omg\mbox{Im}\big[\mb{Tr}\big\{2\big(\mathscr{T}^{(-)}_\alf\wit{N}^\alf(\omg)
+[\mathscr{T}^{(-)}_\alf\wit{N}^\alf(-\omg)]^\dag\big)
+\big(\mathscr{T}^{(+)}_\alf\wit{\rho}(\omg)
+[\mathscr{T}^{(+)}_\alf\wit{\rho}(-\omg)]^\dag\big)\big\}\big],
\end{eqnarray}
where $\wit{N}^\alf(\omg)=\int_0^\infty dtN^\alf(t)e^{i\omg t}$,
and $\wit{\rho}(\omg)=\int_0^\infty dt \rho^{st} e^{i\omg t}$.
Note that we are calculating the noise spectrum of steady state,
thus the stationary density matrix $\rho^{st}$ is used here.
Simply, we have $\wit{\rho}(\omg)=i\rho^{st}/\omg$.
For $\wit{N}^\alf(\omg)$, we can first establish the equation of motion for
$\wit{N}^\alf(t)$, that is
\begin{eqnarray}\label{dNdt}
\frac{d}{dt}N^\alf(t)=-i\mathcal{L}N^\alf(t)-[\mathcal{R}N^\alf(t)-
\mathscr{T}^{(-)}_\alf\rho+\mb{H.c.}].
\end{eqnarray}
Then, from its Fourier transform
\begin{eqnarray}\label{Nomega}
i(\omg-\mathcal{L})\wit{N}^\alf(\omg)&=&\mathcal{R}\wit{N}^\alf(\omg)
+[\mathcal{R}\wit{N}^\alf(-\omg)]^\dag
-\mathscr{T}^{(-)}_\alf\wit{\rho}(\omg)
-[\mathscr{T}^{(-)}_\alf\wit{\rho}(-\omg)]^\dag,
\end{eqnarray}
we can directly obtain $\wit{N}^\alf(\omg)$.

For the charge fluctuations on the central system,
the symmetrized noise spectrum reads
$S_Q(\omg)= \int^{\infty}_{-\infty} d\tau
\la N(\tau)N+NN(\tau) \ra e^{i\omg\tau}]
=4\mb{Re}[\int_0^\infty d\tau S(\tau) e^{i\omg\tau}]$,
where $S(\tau)\equiv\la N(\tau)N \ra=\mb{Tr}\mb{Tr}_B
[U^{\dag}(\tau)NU(\tau)N\rho^{st}\rho_B ]$,
with $U(\tau)=e^{-iH\tau}$ and $N$ the
electron-number operator of the central system.
Using the cyclic property under trace, we have $S(\tau)=\mb{Tr}[N\sgm(\tau)]$,
where $\sgm(\tau)\equiv \mb{Tr}_B[U(\tau)N\rho^{st}\rho_B U^{\dag}(\tau)]$ is
introduced. Noticeably, $\sgm(\tau)$ is nothing but an alternative reduced matrix
which satisfies the same equation of the usual reduced density matrix $\rho(\tau)$,
with the initial condition $\sgm(0)=N\rho^{st}$. Straightforwardly,
its Fourier transform $\wit{\sgm}(\omg)$ can be easily solved from
\begin{eqnarray}\label{rhoomega}
i(\omg-\mathcal{L})\wit{\sgm}(\omg)=\mathcal{R}\wit{\sgm}(\omg)
+[\mathcal{R}\wit{\sgm}(-\omg)]^\dag-N\rho^{st}.
\end{eqnarray}
Then, $S_Q(\omg)=4\mbox{Re}\{ {\rm Tr} [N\wit{\sgm}(\omg)]\}$. More specifically, in most
cases we can carry out this quantity in the eigenstate basis of ``$N$", that is
\begin{eqnarray}\label{SQ}
S_Q(\omg)=4\mbox{Re}\bigg[\sum_kN_k\wit{\sgm}_{k}(\omg)\bigg],
\end{eqnarray}
where $N_k$ is the eigenvalue of $N$ for the eigenstate $|k\ra$, and
$\wit{\sgm}_{k}(\omg)= \la k|\wit{\sgm}(\omg)|k\ra$.

To summarize, we have now constructed a general and quite compact formulation
for quantum transport through mesoscopic systems, particularly for
the calculation of noise spectrum which is usually a difficult problem
by any of other approaches.
Obvious advantages of this formulation include its straightforwardness
and simpleness, together with its applicability to handle the
many-electron correlation, quantum coherence,
and possible inelastic scattering (dissipative)
processes during transport on equal footing.

\section{\label{sec:3}Illustrative Applications}

In this section we demonstrate the above formulation by transport through first
a non-interacting quantum dot, then interacting dot,
and finally two coupled quantum dots.
Remarks with respect to other approaches and completion of
some results which are lacked in existing literature will be made in particular.

\subsection{Non-Interacting Quantum Dot}

For simplicity, we restrict to the simplest case that
only a single dot-state is involved in transport process.
In this example, the electron spin is an irrelevant
degree of freedom which is thus neglected in the description
(its inclusion only needs multiplying an entire factor 2 to the result
obtained in the following).
Accordingly, the dot Hamiltonian reads
$H_{S}=E_0N$, where $N=c^\dag c$ is the number operator.

To proceed, we need to carry out the electron reservoir
(i.e. the electrode) correlation function.
Explicitly, $C^{(\pm)}_{\alf}(t-\tau)=|t_{\alf}|^2\sum_ke^{\pm
iE_k(t-\tau)}f_\alf^{(\pm)}(E_k)$, and its Fourier transform
$C^{(\pm)}_{\alf}(\pm\mathcal{L})=\Gam_{\alf}f_\alf^{(\pm)}(-\mathcal{L})$.
Here, the wide-band approximation for the electrodes is applied,
which results in the {\it energy-independent} coupling constant
$\Gam_{\alf}=2\pi g_\alf|t_{\alf}|^2$, where $g_\alf$ is the density of states.
$f_\alf^{(+)}$ is the usual Fermi function, and $f_\alf^{(-)}\equiv1-f_\alf$.
Since $\mathcal{L}^nc=(-E_0)^n c$, we have
$A_{\alf}^{(\pm)}=C^{(\pm)}_{\alf} (\pm\mathcal{L})c=\Gam_{\alf}f_\alf^{(\pm)}(E_0)c$.
Neglecting the spin degree of freedom, only two states are involved, i.e.,
$|0\ra$ and $|1\ra$ for the empty the occupied states, respectively.
With these identifications, the transport master equation (\ref{QME})
simply reads
\begin{subequations}\label{App1RateEq}
\begin{eqnarray}
\dot{\rho}_{0}&=&-\Gam_{L}\rho_{0}+\Gam_{R}\rho_{1}\label{App1RateEqa},\\
\dot{\rho}_{1}&=&-\Gam_{R}\rho_{1}+\Gam_{L}\rho_{0}\label{App1RateEqb}.
\end{eqnarray}
\end{subequations}
Hereafter, in order to carry out analytic expressions, we assume zero temperature.
By inserting the stationary solution of Eq. (\ref{App1RateEq}) into
the current formula Eq. (\ref{current}),
the well known resonant current is obtained, i.e.,
$\bar{I}_R=-\bar{I}_L=\Gam_{L}\Gam_{R}/(\Gam_{L}+\Gam_{R})$.

Now we turn to calculation of the noise spectrum.
The current fluctuation spectra of the left and right electrodes
follow Eq. (\ref{Salpha}) as
\begin{subequations}\label{App1SLSR}
\begin{eqnarray}
S_L(\omg)&=&2\omg\Gam_{L}\mb{Im}\big[-2\wit{N}^L_{0}(\omg)+\wit{\rho}_{0}(\omg)\big],\\
S_R(\omg)&=&2\omg\Gam_{R}\mb{Im}\big[2\wit{N}^R_{1}(\omg)+\wit{\rho}_{1}(\omg)\big],
\end{eqnarray}
\end{subequations}
where $\wit{\rho}_0(\omg)=i\rho_0^{st}/\omg$, and
$\wit{\rho}_1(\omg)=i\rho_1^{st}/\omg$, with $\rho_0^{st}$ and $\rho_1^{st}$ the
stationary solutions of master equation (\ref{App1RateEq}).
$\wit{N}^L_{0}(\omg)$ and $\wit{N}^R_{1}(\omg)$ can be evaluated
based on Eq. (\ref{Nomega}).
For instance, $\wit{N}^L_{0}(\omg)$ is given by
\begin{subequations}\label{App1Nomg}
\begin{eqnarray}
i\omg\wit{N}^L_{0}(\omg)&=&\Gam_{L}\wit{N}^L_{0}(\omg)-\Gam_{R}\wit{N}^L_{1}(\omg),\\
i\omg\wit{N}^L_{1}(\omg)&=&\Gam_{R}\wit{N}^L_{1}(\omg)
-\Gam_{L}\wit{N}^L_{0}(\omg)+\Gam_{L}\wit{\rho}_{0}(\omg),
\end{eqnarray}
\end{subequations}
and similarly, $\wit{N}^R(\omg)$ can be obtained. Then, straightforwardly, we have
\begin{eqnarray}\label{App1Sw}
S_L(\omg)=S_R(\omg)=2\bar{I}\left[\frac{\Gam_{L}^{2}+\Gam_{R}^2}{\Gam^{2}}
+\frac{2\Gam_{L}\Gam_{R}}{\Gam^{2}}\frac{\omg^{2}}{\Gam^{2}+\omg^{2}}\right],
\end{eqnarray}
where $\bar{I}=\bar{I}_R=-\bar{I}_L$ is the stationary current.
Note that both the zero- and finite-frequency noise spectra of the two
tunneling currents are identical for this simple system.

To calculate the charge fluctuation spectrum on the central quantum dot, based on Eq.
(\ref{SQ}) we simply have $S_Q(\omg)=4\mbox{Re} [\wit{\sgm}_{1}(\omg)]$, where the fact
that $N_0=0$ and $N_1=1$ have been taken into account.
Following Eq. (\ref{rhoomega}), $\wit{\sgm}_{1}(\omg)$ is given by the following coupled
equations
\begin{subequations}\label{App1rhoomg}
\begin{eqnarray}
i\omg\wit{\sgm}_{0}(\omg)&=&\Gam_{L}\wit{\sgm}_{0}(\omg)
-\Gam_{R}\wit{\sgm}_{1}(\omg),\\
i\omg\wit{\sgm}_{1}(\omg)&=&\Gam_{R}\wit{\sgm}_{1}(\omg)
-\Gam_{L}\wit{\sgm}_{0}(\omg)-\rho_{1}^{st}.
\end{eqnarray}
\end{subequations}
Accordingly, we obtain
$S_Q(\omg)=4\bar{I}/(\Gam^{2}+\omg^{2})$.
Combining all the three components of noise spectrum according to \Eq{Sc}
and assuming a symmetric configuration (i.e. $a=b=1/2$), we obtain
\begin{eqnarray}\label{App1SC}
S(\omg)=\bar{I}\left[1+\left(1-\frac{4\Gam_L\Gam_R}{\Gam^2}\right)
\frac{\Gam^2}{\Gam^2+\omg^2}\right].
\end{eqnarray}
This is the well-known result for resonant tunneling
through symmetric double-barrier structures,
which was obtained first
by Chen and Ting based on the non-equilibrium path-integral technique \cite{Ch-T},
and also by B\"{u}ttiker by using the scattering approach \cite{Butti},
where the so-called Fano factor $F(\omg)=S(\omg)/2\bar{I}$,
especially its value at zero frequency
$F(0)=1-2\Gam_{L}\Gam_{R}/\Gam^{2}$, was discussed in particular.
Here, we presented an alternative derivation which seems elegant and interesting.
More importantly, its generalization to more complicated systems
is straightforward, as we are going to illustrate soon.

\subsection{Interacting QD with Zeeman Splitting}

We now consider transport through an interacting quantum dot described by
\begin{eqnarray}
H_{S}= \sum _{\sgm}E_{\sgm}N_{\sgm}+UN_{\upa}N_{\dwa},
\end{eqnarray}
where $E_\dwa$ and $E_\upa$ are the electron levels with finite Zeeman splitting,
and $N_\sgm=c^\dag_\sgm c_\sgm$ is the electron number operator for spin $\sgm$
(it should not be confused with the reduced operator $\sgm(\tau)$ introduced
in the calculation of $S_Q(\omega)$ ).
In this model, the $U$-term accounts for the many-body Coulomb correlation.
As in the previous example, we should first carry out the spectral function
$C^{(\pm)}_{\alf\sgm\sgm'}$, and the operator $A_{\alf\sgm}^{(\pm)}$.
Since the interaction does not flip the electron spin, the spectral function
must be diagonal with respect to the spin indices, {\it i.e.},
$C^{(\pm)}_{\alf\sgm\sgm'}=\delta_{\sgm\sgm'}C^{(\pm)}_{\alf\sgm\sgm}$.
Still under the wide-band approximation, we have
$A_{\alf\sgm}^{(\pm)}=\Gam_{\alf}f_{\alf}^{(\pm)}
(W_{\sgm})c_{\sgm}$,
where $W_{\sgm}\equiv
E_{\upa}\delta_{\upa\sgm}+ E_{\dwa}\delta_{\dwa\sgm}+ U(n_{\upa}\delta_{\dwa\sgm}
+n_{\dwa}\delta_{\upa\sgm})$, which sets up four regimes for the applied bias voltage,
i.e., (i) $\mu_L>E_\dwa>\mu_R$,
(ii) $\mu_L>E_{\upa},E_{\dwa}>\mu_R$,
(iii) $\mu_L\!>\!E_{\dwa}+U,E_{\upa},E_{\dwa}\!>\!\mu_R$, and
(iv) $\mu_L\!\!>\!E_{\upa}\!+\!U,E_{\dwa}\!+\!U,E_{\upa},E_{\dwa}\!>\!\mu_R$.

For the sake of being compact, here we only present detailed derivation for regime (ii).
In this regime, the involved states include
 $|0\ra$ (empty dot), $|\!\upa\,\ra$ (occupation by a spin-up electron),
and $|\!\dwa\,\ra$ (occupation by a spin-down electron).
Accordingly, the transport master equation reads
\begin{eqnarray}\label{app2-2QME}
\dot{\rho}_{0}&=&-2\Gam_{L}\rho_{0}+\Gam_{R}(\rho_{\dwa}+\rho_{\upa}),\nl
\dot{\rho}_{\dwa}&=&-\Gam_{R}\rho_{\dwa}+\Gam_{L}\rho_{0}, \\
\dot{\rho}_{\upa}&=&-\Gam_{R}\rho_{\upa}+\Gam_{L}\rho_{0}.\nonumber
\end{eqnarray}
Its stationary solution gives rise to the steady-state current
$\bar{I}=\bar{I}_R=-\bar{I}_L=2\Gam_{L}\Gam_{R}/(2\Gam_{L}+\Gam_{R})$.

In the above mentioned representation of occupation states, $S_{L/R}(\omg)$
are expressed as
\begin{subequations}\label{App2-2SLSR}
\begin{eqnarray}
S_L(\omg)&=&4\omg\Gam_{L}\mb{Im}\big[-2\wit{N}^L_{0}(\omg)+\wit{\rho}_{0}(\omg)\big],\\
S_R(\omg)&=&2\omg\Gam_R\mb{Im}\big[2\wit{N}^R_\upa(\omg)+2\wit{N}^R_\dwa(\omg)
+\wit{\rho}_{\upa}(\omg)+\wit{\rho}_{\dwa}(\omg)\big].
\end{eqnarray}
\end{subequations}
$\wit{N}^{L}_{0}(\omg)$, $\wit{N}^R_\upa(\omg)$ and $\wit{N}^R_\dwa(\omg)$
can be obtained from Eq. (\ref{Nomega}).
For $\wit{N}^{L}_{0}(\omg)$, for instance, the relevant equations explicitly read
\begin{eqnarray}\label{App2-2dNL}
i\omg\wit{N}^L_{0}(\omg)&=&2\Gam_{L}\wit{N}^L_{0}(\omg)
-\Gam_{R}\wit{N}^L_{\upa}(\omg)-\Gam_{R}\wit{N}^L_{\dwa}(\omg),\nl
i\omg\wit{N}^L_{\upa}(\omg)&=&\Gam_{R}\wit{N}^L_{\upa}(\omg)
-\Gam_{L}\wit{N}^L_{0}(\omg)+\Gam_{L}\wit{\rho}_{0}(\omg),   \\
i\omg\wit{N}^L_{\dwa}(\omg)&=&\Gam_{R}\wit{N}^L_{\dwa}(\omg)
-\Gam_{L}\wit{N}^L_{0}(\omg)+\Gam_{L}\wit{\rho}_{0}(\omg) \nonumber.
\end{eqnarray}
Note that, in the above equations,
$\wit{\rho}(\omg)=i\rho^{st}/\omg$, and $\rho^{st}$ is the stationary
solution of \Eq{app2-2QME}. Straightforwardly, we obtain
\begin{eqnarray}\label{App2Sw}
S_L(\omg)=S_R(\omg)=2\bar{I}\left[\frac{(2\Gam_L)^2+\Gam_{R}^2}{(2\Gam_L+\Gam_R)^2}
+\frac{2(2\Gam_{L})\Gam_{R}}{(2\Gam_L+\Gam_R)^2}\frac{\omg^{2}}
{(2\Gam_L+\Gam_R)^2+\omg^{2}}\right].
\end{eqnarray}
Also, in the same occupation-state representation, $S_Q(\omg)$ simply reads
$S_Q(\omg)=4\mb{Re}[\wit{\sgm}_{\upa}(\omg)+\wit{\sgm}_{\dwa}(\omg)]$, where
$\wit{\sgm}_{\upa} (\omg)$ and $\wit{\sgm}_{\dwa}(\omg)$ can be solved from
\begin{eqnarray}\label{App2-2rhoomega}
i\omg\wit{\sgm}_0(\omg)&=&2\Gam_L\wit{\sgm}_0(\omg)-\Gam_R\wit{\sgm}_\upa(\omg)
-\Gam_R\wit{\sgm}_\dwa(\omg),\nl
i\omg\wit{\sgm}_\upa(\omg)&=&\Gam_R\wit{\sgm}_\upa(\omg)-\Gam_L\wit{\sgm}_{0}(\omg)
-\rho_\upa^{st},  \\
i\omg\wit{\sgm}_\dwa(\omg)&=&\Gam_R\wit{\sgm}_\dwa(\omg)-\Gam_L\wit{\sgm}_{0}(\omg)
-\rho_\dwa^{st}.\nonumber
\end{eqnarray}
Its solution gives
\begin{eqnarray}\label{App2-2SQ}
S_Q(\omg)&=&4\bar{I}\left[\frac{1}{(2\Gam_L+\Gam_R)^2+\omg^2}\right].
\end{eqnarray}
In Table I and II all the components of the noise spectrum are listed, together with those
of other voltage regimes, in which the derivation is completely the same as above.

\begin{table}
\caption{\label{Table1}Zero-frequency and frequency-dependent noise components
in the different bias regimes (i), (ii) and (iv),
where $\Gam=\Gam_L+\Gam_R$ is the total level-broadening width. }
\begin{tabular}{cccc}\hline\hline
bias regime &(i)&(ii)&(iv)
\\ \hline
\pb[c][1cm]{1.5cm}{\be S(0)\ee}
&\pb[c][1cm]{3.5cm}{\be\frac{2\Gam_L\Gam_R(\Gam_L^2+\Gam_R^2)}{\Gam^3}\ee}
&\pb[c][1cm]{4.5cm}{\be\frac{4\Gam_L\Gam_R(4\Gam_L^2+\Gam_R^2)}{(2\Gam_L+\Gam_R)^3}\ee}
&\pb[c][1cm]{3.5cm}{\be\frac{4\Gam_L\Gam_R(\Gam_L^2+\Gam_R^2)}{\Gam^3}\ee}
\\
\pb[c][1cm]{1.5cm}{\be S_L(\omg) \ee}
&\pb[c][1cm]{3.5cm}{\be\frac{4\Gam_L^2\Gam_R^2}{\Gam^3}\frac{\omg^2}{\Gam^2+\omg^2}\ee}
&\pb[c][1cm]{4.5cm}{\be\frac{16\Gam_L^2\Gam_R^2}{(2\Gam_L+\Gam_R)^{3}}\frac{\omg^2}{(2\Gam_L+\Gam_R)^2+\omg^2}\ee}
&\pb[c][1cm]{3.5cm}{\be\frac{8\Gam_L^2\Gam_R^2}{\Gam^3}\frac{\omg^2}{\Gam^2+\omg^2}\ee}
\\
\pb[c][1cm]{1.5cm}{\be S_R(\omg) \ee}
&\pb[c][1cm]{3.5cm}{\be\frac{4\Gam_L^2\Gam_R^2}{\Gam^3}\frac{\omg^2}{\Gam^2+\omg^2}\ee}
&\pb[c][1cm]{4.5cm}{\be\frac{16\Gam_L^2\Gam_R^2}{(2\Gam_L+\Gam_R)^{3}}\frac{\omg^2}{(2\Gam_L+\Gam_R)^2+\omg^2}\ee}
&\pb[c][1cm]{3.5cm}{\be\frac{8\Gam_L^2\Gam_R^2}{\Gam^3}\frac{\omg^2}{\Gam^2+\omg^2}\ee}
\\
\pb[c][1cm]{1.5cm}{\be \omg^2S_Q(\omg) \ee}
&\pb[c][1cm]{3.5cm}{\be\frac{4\Gam_L\Gam_R}{\Gam}\frac{\omg^2}{\Gam^2+\omg^2}\ee}
&\pb[c][1cm]{4.5cm}{\be\frac{8\Gam_L\Gam_R}{(2\Gam_L+\Gam_R)}\frac{\omg^2}{(2\Gam_L+\Gam_R)^2+\omg^2}\ee}
&\pb[c][1cm]{3.5cm}{\be\frac{8\Gam_L\Gam_R}{\Gam}\frac{\omg^2}{\Gam^2+\omg^2}\ee}
\\ \hline\hline
\end{tabular}
\end{table}

\begin{table}
\caption{\label{Table2}Zero-frequency and frequency-dependent noise components
in regimes (iii),
where $\Gam=\Gam_L+\Gam_R$ is the total level-broadening width. }
\begin{tabular}{cc}\hline\hline
bias regime &(iii)
\\ \hline
\pb[c][1cm]{1.5cm}{\be S(0)\ee}
&\pb[c][1cm]{5cm}{\be\frac{2\Gam_L\Gam_R(\Gam_L+2\Gam_R)}{\Gam^5}(\Gam^3-3\Gam_L\Gam_R^2)\ee}
\\
\pb[c][1cm]{1.5cm}{\be S_L(\omg)\ee} &\pb[c][1cm]{5cm}{
\begin{align*}
\frac{2\Gam_L^2\Gam_{R}(\Gam^3-\Gam_L^2\Gam_R+\Gam_L\Gam_R^2+5\Gam_R^3)\omg^2}{\Gam^5(\Gam^2+\omg^2)}
+\frac{4\Gam_L^3\Gam_R(\Gam_L^2-2\Gam_R^2)\omg^2}{\Gam^3(\Gam^2+\omg^2)^2}
\end{align*}}
\\
\pb[c][1cm]{1.5cm}{\be S_R(\omg)\ee} &\pb[c][1cm]{5cm}{
\begin{align*}
\frac{6\Gam_L^2\Gam_R^3(\Gam_L+2\Gam_R)}{\Gam^5}\frac{\omg^{2}}{\Gam^2+\omg^2}
-\frac{4\Gam_L^3\Gam_R^2(2\Gam_L+3\Gam_R)}{\Gam^3}\frac{\omg^2}{(\Gam^2+\omg^2)^2}
\end{align*}}
\\
\pb[c][1cm]{1.5cm}{\be \omg^2S_Q\ee} &\pb[c][1cm]{5cm}{
\begin{align*}
\frac{2\Gam_L\Gam_R(\Gam^2+3\Gam_R^2)}{\Gam^3}\frac{\omg^2}{\Gam^2+\omg^2}
+\frac{4\Gam_L^2\Gam_R(\Gam_L+2\Gam_R)}{\Gam^3}\frac{\omg^4}{(\Gam^2+\omg^2)^2}
\end{align*}}
\\ \hline\hline
\end{tabular}
\end{table}


In Ref.\ \onlinecite{Schon03} the zero-frequency noise was carried out
for the same system of this example, based on a complicated non-equilibrium
real time diagrammatic technique.
Here we extended the result to the whole frequency regime, using a much
simpler and more straightforward approach.
We notice that the noise spectra in regimes (i) and (iv) differ from each
other only by an overall factor 2.
The reason is that the electron near the Fermi surface of the left electrode
can only pass through the spin-down level of the quantum dot in regime (i),
but can freely pass through the two levels in regime (iv) due to
completely overcoming the Coulomb blockade.
As a consequence, the result in regime (iv) is also the same as that of the
non-interacting dot obtained in the previous subsection.
In the Coulomb blockade regime (ii), electron can only
pass through either the spin-down or the spin-up level in an exclusive manner.
Compared to the situation of regime (i), the entrance probability of electron
from the left electrode to the dot is enhanced by a factor of 2,
whereas the leaving probability to the right electrode is the same.
Therefore, in all the components of noise spectrum in regime (ii),
``$2\Gamma_L$" replaces the ``$\Gamma_L$" in the result of regime (i).
In regime (iii), the Coulomb blockade is partially overcome, i.e.,
electron can enter the dot even there has been already an electron
on the spin-up state in the quantum dot.
The noise spectrum in this regime is relatively complicated, and
seems beyond an intuitive simple interpretation.
However, the unique features of noise spectrum in regime (ii) and (iii)
may provide a pathway to distinguish Coulomb blockade phenomena from
that resulting from non-interacting multi-levels,
as emphasized in Ref.\ \onlinecite{Schon03} based on the zero-frequency noise.
Moreover, the ``asymmetry" of $\Gamma_L$ and $\Gamma_R$ may provide
a useful tool to determine their respective values, which is very important
in molecular electronics, due to the need to identify the subtle connection
of the molecule with the electrodes.


\subsection{Two Coupled Quantum Dots}

In this subsection we study the noise characteristics of transport through
a pair of coupled quantum dots, as shown in Fig.\ 2.
To highlight the underlying coherence effect, which surely goes beyond the scope of
the {\it classical} rate equation approach, we would like to neglect the many-body
Coulomb correlation. (Inclusion of it is straightforward but will make the solution
more complicated). Accordingly, the system Hamiltonian reads
\begin{eqnarray}
\hat{H}_{S} =
E_{1}c^{\dag}_{1}c_{1}+E_{2}c^{\dag}_{2}c_{2}+\Omg(c_{1}^{\dag}c_{2}+c_{2}^{\dag}c_{1}) .
\end{eqnarray}
To analytically carry out the operators $A_{\alpha}^{(\pm)}$, it will be useful
to diagonalize this Hamiltonian. To this end, we introduce
a pair of new electronic operators,
$\tilde{c}_{1}=uc_{1}+vc_{2}$ and $\tilde{c}_{2}=uc_{2}-vc_{1}$.
The desired Hamiltonian
$H_{S}=\vpl_{1} \tilde{c}^{\dag}_{1}\tilde{c}_{1}+\vpl_{2}\tilde{c}^{\dag}_{2}\tilde{c}_{2}$,
gives rise to the diagonalization condition $(E_{2}-E_{1})uv+\Omg(u^{2}+v^{2})=0$.
Together with $ u^{2}+v^{2}=1$, we can obtain
the values of $u$ and $v$,  and also the eigen-energies
$\vpl_{1} =E_{1}u^{2}+E_{2}v^{2}+2\Omg uv$, and
$\vpl_{2}=E_{1}v^{2}+E_{2}u^{2}-2\Omg uv$.

\begin{figure}
\begin{center}
\includegraphics*[width=7cm,height=5cm,keepaspectratio]{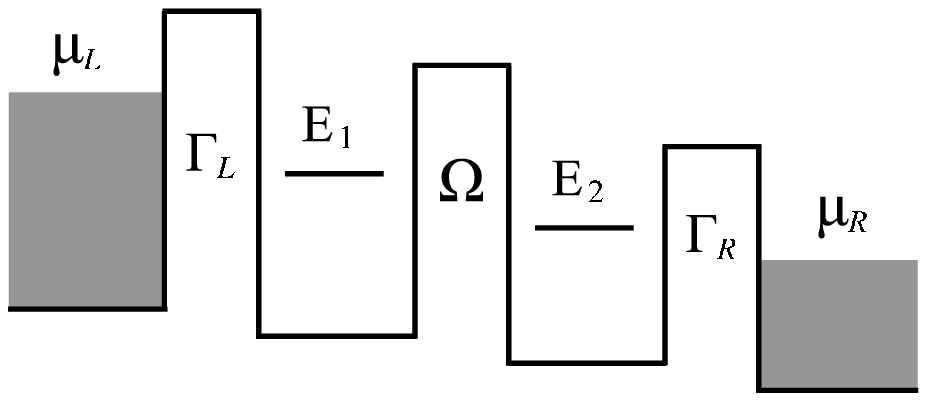}
\caption{Schematic illustration for transport through a pair of coupled quantum dots.}
\end{center}
\end{figure}

As done previously, under the wide-band approximation,
we can first carry out the spectral function $C^{(\pm)}_{\alpha}$, then the operators
$A_{L}^{(\pm)}=\Gam_L[uf^{(\pm)}_{L}(\vpl_1)\tilde{c}_1-vf^{(\pm)}_L(\vpl_2)\tilde{c}_2]$,
and $A_{R}^{(\pm)}=\Gam_R[vf^{(\pm)}_{R}(\vpl_1)\tilde{c}_1+uf^{(\pm)}_R(\vpl_2)\tilde{c}_2]$.
Neglecting the irrelevant spin degree of freedom, the four relevant states of the double
dots are  $|0\ra$ (empty dots), $|1\ra$ (occupation of the left dot),
$|2\ra$ (occupation of the right dot), and $|d\ra$ (occupation of both dots).
In this state representation, Eq. (\ref{QME}) becomes
\begin{eqnarray}\label{App3RateEq}
\dot{\rho}_0&=&\Gam_R\rho_2-\Gam_L\rho_0,\nl
\dot{\rho}_1&=&\Gam_L\rho_0+\Gam_R\rho_d+i\Omg(\rho_{12}-\rho_{21}),\nl
\dot{\rho}_2&=&-(\Gam_L+\Gam_R)\rho_{2}-i\Omg(\rho_{12}-\rho_{21}), \\
\dot{\rho}_d&=&\Gam_L\rho_2-\Gam_R\rho_d,\nl
\dot{\rho}_{12}&=&i\Delta\rho_{12}-i\Omg(\rho_1-\rho_2)-\frac{1}{2}(\Gam_L
+\Gam_{R})\rho_{12},\nonumber
\end{eqnarray}
where $\rho_{21}=\rho_{12}^\ast$, and $\Delta\equiv E_2-E_1$.
Note that the off-diagonal density matrix element $\rho_{12}$ purely
arises from the quantum coherence between the two coupled dots,
and has has no classical counterpart.
From the stationary solution of \Eq{App3RateEq}, we can calculate the steady-state current as
\begin{eqnarray}
\bar{I}_R=\frac{4\Gam_L\Gam_R(\Gam_L+\Gam_R)\Omg^2}
{\Gam_L\Gam_R[(\Gam_L+\Gam_R)^2+4\Delta^2]+4(\Gam_L+\Gam_R)^2\Omg^2} .
\end{eqnarray}
This result was also obtained in Ref.\ \onlinecite{Gurv96}.

In the occupation-state representation, we express the electrode current fluctuation
spectra as
\begin{subequations}\label{App3SLSR}
\begin{eqnarray}
S_L(\omg)&=&2\omg\Gam_L\mb{Im}\big[-2\wit{N}^L_0(\omg)
-2\wit{N}^L_2(\omg)+\wit{\rho}_0(\omg)+\wit{\rho}_2(\omg)\big]\label{App3SLSRa},\\
S_R(\omg)&=&2\omg\Gam_R\mb{Im}\big[2\wit{N}^R_2(\omg)
+2\wit{N}^R_d(\omg)+\wit{\rho}_2(\omg)+\wit{\rho}_d(\omg)\big],
\end{eqnarray}
\end{subequations}
where $\wit{\rho}(\omg)=i\rho^{st}/\omg$, given by the stationary solution
of \Eq{App3RateEq}. For $S_L(\omg)$, $\wit{N}^L_0(\omg)$ and $\wit{N}^L_2(\omg)$ are
solved from
\begin{eqnarray}\label{App3dNL}
i\omg\wit{N}^L_0(\omg)&=&-\Gam_R\wit{N}^L_2(\omg)+\Gam_L\wit{N}^L_0(\omg),\nl
i\omg\wit{N}^L_1(\omg)&=&-\Gam_L\wit{N}^L_0(\omg)-\Gam_R\wit{N}^L_d(\omg)
-i\Omg[\wit{N}^L_{12}(\omg)-\wit{N}^L_{21}(\omg)]+\Gam_L\wit{\rho}_0(\omg),\nl
i\omg\wit{N}^L_2(\omg)&=&(\Gam_L+\Gam_R)\wit{N}^L_2(\omg)
+i\Omg[\wit{N}^L_{12}(\omg)-\wit{N}^L_{21}(\omg)],    \\
i\omg\wit{N}^L_d(\omg)&=&-\Gam_L\wit{N}^L_2(\omg)
+\Gam_R\wit{N}^L_d(\omg)+\Gam_L\wit{\rho}_2(\omg),\nl
i\omg\wit{N}^L_{12}(\omg)&=&-i\Omg[\wit{N}^L_1(\omg)-\wit{N}^L_2(\omg)]
+(\Gam_L+\Gam_R)\wit{N}_{12}^L(\omg)/2.\nonumber
\end{eqnarray}
In the following we would like to present analytic result for $\Delta=0$,
and numerical result for $\Delta\neq 0$.
Also, for simplicity, we assume $\Gam_L=\Gam_R=\Gam$.
Straightforwardly, based on the solution of \Eq{App3dNL}, we obtain
$S_L(\omg)=S_L(0)+S'_L(\omg)+S''_L(\omg)$, where
\begin{subequations}
\begin{eqnarray}\label{App3S0}
S_L(0)=\frac{4\Gam\Omg^2(\Gam^{4}-2\Gam^2\Omg^2+8\Omg^4)}{(\Gam^2+4\Omg^2)^{3}},
\end{eqnarray}
\begin{eqnarray}\label{App3S1}
S'_L(\omg)=\frac{2\Gam\Omg^2}{\Gam^2+4\Omg^2}\frac{\omg^2}{\Gam^2+\omg^2},
\end{eqnarray}
\begin{eqnarray}\label{App3S2}
S''_L(\omg)&=&\frac{2\Gam^3\Omg^3(3\Gam^2-4\Omg^2)}{(\Gam^2+4\Omg^2)^3}
\bigg[\frac{\omg}{\Gam^2+(\omg-2\Omg)^2}-\frac{\omg}{\Gam^2+(\omg+2\Omg)^2}\bigg]\nl
&&\!\!+\frac{\Gam^3\Omg^2(12\Omg^2-\Gam^2)}{(\Gam^2+4\Omg^2)^3}
\bigg[\frac{(\omg-2\Omg)\omg}{\Gam^2+(\omg-2\Omg)^2}
+\frac{(\omg+2\Omg)\omg}{\Gam^2+(\omg+2\Omg)^2}\bigg].
\end{eqnarray}
\end{subequations}
Note that $S'_L(\omg)$ is the well-known Lorentzian which stands for the
incoherent component with similar structure as those shown in Table I and II,
whereas $S''_L(\omg)$ stems from the coherent coupling between the two quantum dots.
For $S_R(\omg)$, precisely the same result can be carried out by the same procedures.
For the charge fluctuations on the double dots, we have
$S_Q(\omg)=4\mb{Re}[\wit{\sgm}_1(\omg)+\wit{\sgm}_2(\omg)+2\wit{\sgm}_d(\omg)]$, where
$\wit{\sgm}(\omg)$ is obtained from
\begin{eqnarray}\label{App3rhoomega}
i\omg\wit{\sgm}_0(\omg)&=&-\Gam_R\wit{\sgm}_2(\omg)+\Gam_L\wit{\sgm}_0(\omg),\nl
i\omg\wit{\sgm}_1(\omg)&=&-\Gam_L\wit{\sgm}_0(\omg)
-\Gam_R\wit{\sgm}_d(\omg)-i\Omg[\wit{\sgm}_{12}(\omg)
-\wit{\sgm}_{21}(\omg)]-\rho_1^{st},\nl
i\omg\wit{\sgm}_2(\omg)&=&(\Gam_L+\Gam_R)\wit{\sgm}_2(\omg)
+i\Omg[\wit{\sgm}_{12}(\omg)-\wit{\sgm}_{21}(\omg)]-\rho_2^{st},   \\
i\omg\wit{\sgm}_d(\omg)&=&-\Gam_L\wit{\sgm}_2(\omg)+\Gam_R\wit{\sgm}_d(\omg)-2\rho_d^{st},\nl
i\omg\wit{\sgm}_{12}(\omg)&=&-i\Omg[\wit{\sgm}_1(\omg)
-\wit{\sgm}_2(\omg)]+(\Gam_L+\Gam_R)\wit{\sgm}_{12}(\omg)/2. \nonumber
\end{eqnarray}
Then, we immediately arrive at
\begin{eqnarray}\label{App3SQ}
S_{Q}(\omg)=\frac{8\Gam\Omg^2}{(\Gam^2+4\Omg^2)(\Gam^2+\omg^2)}.
\end{eqnarray}

\begin{figure}
\begin{center}
\includegraphics*[width=8cm,height=8cm,keepaspectratio]{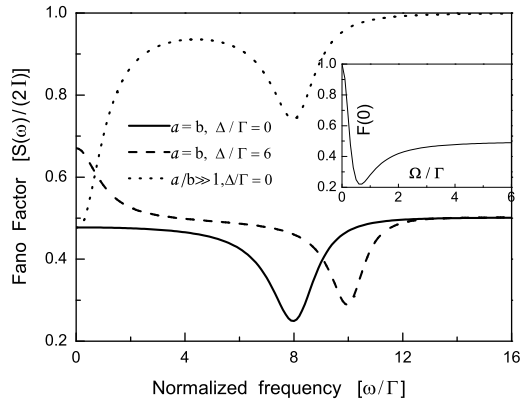}
\caption{
Fano factor $F(\omg)$ versus the normalized frequency $\omg/\Gam$.
The signature of quantum coherence between the two coupled quantum dots
is the dip appeared at the Rabi frequency.
In the numerical calculation, we assume $\Omg=4\Gam$.
Other geometric parameters are labelled in the figure,
and the corresponding results are shown by the solid, dashed, and dotted
curves, respectively.
Inset: zero frequency Fano factor, which
shows a non-monotonic dependence on the scaled coupling strength $\Omega/\Gamma$. }
\end{center}
\end{figure}

The noise characteristics is numerically shown in Fig.\ 3.
In addition to Ref.\ \onlinecite{Milb}, some more discussions
are briefly presented as follows:
(i) Due to the coherent coupling between the two quantum dots,
before the transport electron arrives at the right electrode, it may stay
in the coupled dots for some time and oscillate between them.
During this Rabi oscillation process, it prevents other electrons
from entering the dots. As a result, a ``dip" appears in the noise spectrum
at the Rabi frequency. For example, for a setup parameterized by $\Omg=4\Gam$,
$\Delta=0$ and $a=b=1/2$, the dip locates at $\omg_R=2\Omg$, as shown by the
solid curve in Fig.\ 3.
This interpretation is illustrated further by the result from
a pair of off-resonantly coupled quantum dots. Since for this case the
oscillation frequency is $\omg'_R=\sqrt{\Delta^2 +(2\Omg)^2}$,
the ``dip" should move to higher frequency. This is shown by the dashed curve
in Fig.\ 3 where $\Delta=6\Gam$ is assumed.
(ii) In our approach, we explicitly distinguish the three components
contributing to the noise spectrum, which enable us to gain deeper insight.
For example, to demonstrate the effect of charge-number fluctuations on
the coupled dots, we consider a very asymmetric configuration ($a/b\gg 1$).
Noting that $S_L(\omega)=S_R(\omega)$ \cite{note-1}, from \Eq{Sc} we
conclude that the charge-number fluctuation noise $S_Q(\omega)$
would play negligible role on the entire noise spectrum if $a/b\gg 1$.
In particular, the difference of the noise spectra of
$a=b$ and  $a/b\gg 1$ can give important information for $S_Q(\omega)$.
A simple comparison between the solid and dotted curves in Fig.\ 3
indicates that $S_Q(\omega)$ would reduce the entire noise spectrum.
Also, for $a/b\gg 1$, at high frequency limit the noise spectrum
approaches to unity, which is consistent with the result of transport
through a quantum point contact \cite{Butti}.
(iii) Concerning the zero frequency noise [c.f. \Eq{App3S0}],
it shows a non-monotonic dependence on the (scaled) Rabit coupling
strength $\Omg/\Gam$. In the inset of Fig. 3, we plot the corresponding
Fano factor $F(0)\equiv S(0)/2\bar{I}$.
Interestingly, it will be maximally suppressed at $\sim 0.65\Omg/\Gam$.
By increasing the coupling strength ($\Omg\gg\Gam$),
the zero-frequency Fano factor approaches $1/2$, behaving similarly as
the symmetric double-barrier resonant tunneling structure \cite{Butti}.
In the opposite limit, $\Omg\ll\Gam$, the Fano factor reaches unity
which corresponds to the Schottky-type noise.

Finally, we make a brief technical comment. As we have mentioned,
this interesting system has been studied in Ref.\ \onlinecite{Milb},
where the quantum trajectory approach was employed and particular
attention was focused on the effect of quantum coherence.
It was also proposed there that the {\it quantum trajectory approach} provides
a {\it powerful tool} to calculate transport noise spectrum in the presence
of {\it internal} quantum coherence in the transport systems.
In this subsection, we have shown that the ``$n$"-resolved master equation
approach can easily solve those types of problems.
Moreover, we notice that the quantum trajectory approach has severe limitation
since it is based on the unravelling of Lindblad-type master equation.
We believe that in many transport systems the simple Lindblad-type master
equation cannot be obtained. A simple example is the one studied in the
previous subsection. In this case, the quantum trajectory approach proposed
in Ref.\ \onlinecite{Milb} would fail.
On the contrary, the ``$n$"-resolved master equation approach can efficiently
solve the problems in the presence of both quantum coherence and many-body
Coulomb interactions. An interesting example may be the one studied
in this subsection by including further the Coulomb interactions, which is
currently under our study.

\section{Conclusion}\label{sectionconclusion}

To summarize, based on the ``$n$"-resolved quantum master equation
we have presented an efficient approach to calculating
of the current noise spectrum.
It overcomes the drawbacks of the classical rate equation methods
and the recently proposed quantum trajectory approach,
since the former cannot
account for quantum coherence, whereas the latter seems unable to handle
many-body Coulomb correlations.
In practice, this approach is much simpler
and more straightforward than other well-known
full quantum mechanical methods such as the Landauer-B\"uttiker
scattering matrix theory and the nonequilibrium Green's function technique.
These advantages have been preliminarily illustrated by a number of examples
of quantum transport through quantum dots, where the known results
and new insights are obtained in a unified and transparent way.
Further application to more complicated cases is worthwhile and straightforward,
which will be the topic of our forthcoming work.


\begin{acknowledgments}
Support from the National Natural Science Foundation of China
under No.\ 60376037, 60425412 and 90503013,
the Major State Basic Research Project No.\ G001CB3095 of China,
and the Research Grants Council of the Hong
Kong Government are gratefully acknowledged.
\end{acknowledgments}



\clearpage


\begin{thebibliography}{61}
\bibitem{Abb}
\bibinfo{author}{\bibnamefont{{Since the transport circuit
is closed, extra electrons arrived at the collector will flow back
into the emitter via the external circuit. This rapid relaxation
processes quickly bring the reservoirs to the local thermal
equilibrium state}}}.
\bibitem{note-1}
We have proved this property for the system of coupled quantum dots,
which also holds for many other (but not all) systems.
\end{thebibliography}
\end{document}